\documentclass[conference]{IEEEtran}
\usepackage[T1]{fontenc}
\usepackage{amsmath,amssymb}
\usepackage{booktabs,array,tabularx}
\usepackage{graphicx,xcolor}
\usepackage{algorithm,algorithmic}
\usepackage{listings}
\usepackage{tikz}
\usetikzlibrary{arrows.meta,positioning,fit,backgrounds}
\usepackage{cite}
\usepackage[hidelinks]{hyperref}
\usepackage{url}
\usepackage{microtype}
\newcolumntype{Y}{>{\raggedright\arraybackslash}X}
\begin{document}
\title{Evolving Inspectable O-RAN Slicing xApps with LLMs}
\author{\IEEEauthorblockN{Faezeh Dehghan Tarzjani, Bhaskar Krishnamachari}
\IEEEauthorblockA{Department of Electrical and Computer Engineering, University of Southern California\\
\{dehghant, bkrishna\}@usc.edu}}
\maketitle

\begin{abstract}
Open RAN (O-RAN) slicing xApps must adapt resource allocations to changing channel conditions and traffic demands while meeting service-level agreements (SLAs). Deep reinforcement learning can produce adaptive policies, but their allocation rules remain encoded in neural-network parameters. Our goal is to retain this adaptability while making the controller’s decision logic directly inspectable and editable by operators. We use a large language model (LLM) to evolve slicing controllers as compact Python programs whose decision logic remains readable and editable after optimization. The LLM proposes and revises candidates offline, while a calibrated simulator scores them, and the selected decision module runs unchanged in the O-RAN control path. On the NSF POWDER 5G testbed, the evolved controller releases resources from a guaranteed slice whose throughput target becomes unattainable under a sustained channel fade, improving best-effort throughput from 158.2 to 228.6~Mbps, a 44.5\% gain over the best static allocation. Since the controllers are readable source code, their behavior can be predicted from their equations, defects can be diagnosed by reading the code, and calibration errors can be corrected with one-line edits, reducing SLA misses from 79.9\% to 2.2\% in one case and more than doubling fitness in another. In a four-slice trace-driven simulation calibrated to the same testbed, evolutionary search achieves higher average evaluation scores than independent prompting at a matched proposal budget, with mean normalized gains on held-out traces of 16.3\% for prompting alone, 32.1\% for evolution from scratch, and 51.0\% for evolution from a starting program.
\end{abstract}


\section{Introduction}
\label{sec:introduction}
Open RAN exposes radio access network control through programmable components and open interfaces. The near-real-time RAN Intelligent Controller (near-RT RIC) hosts xApps that observe radio measurements and issue control decisions~\cite{oran-arch}. A slicing xApp divides a shared pool of physical resource blocks (PRBs) among services with diverse Quality of Service (QoS) and service-level agreement (SLA) requirements (such as throughput, latency, or reliability guarantees). An effective slicing controller must track both changing traffic demand and channel quality, particularly when channel quality drops far enough that a slice's throughput target becomes physically unattainable.

A sustained channel fade can make a slice's throughput target
unattainable even at its maximum PRB share, a regime we call
a deep null. Our scoring rule rewards meeting the throughput
requirement and penalizes a miss, without giving partial
credit for falling short. Releasing resources above the
slice's minimum share can therefore benefit best-effort
traffic. When multiple guaranteed slices compete for the
same pool, the controller must also decide which targets
to fund and when to restore service as channels recover.

Writing and tuning rules that handle these decisions correctly is difficult. Deep reinforcement learning (DRL) has been applied to O-RAN slicing under changing traffic and resource contention~\cite{dorcheh2025dora,yan2025near,tsampazi2024pandora,polese2022colo}, and post-hoc explainability frameworks can interpret the resulting policies~\cite{fiandrino2023explora}, but the policy's decision rules remain encoded in
learned parameters.
We want policies whose allocation rules and state updates live in source code that an operator can open, trace, and modify — what we call \emph{inspectable} policies. To produce them automatically, we use a large language model (LLM) to generate, evaluate, and select slicing controllers as compact Python programs through iterative evolutionary search. Prompting an LLM to write a controller in a single pass does not reliably produce correct release logic (Sections~\ref{subsec:search-stats} and~\ref{subsec:predictions}).

\begin{figure}[t]
\centering
\includegraphics[width=0.85\columnwidth]{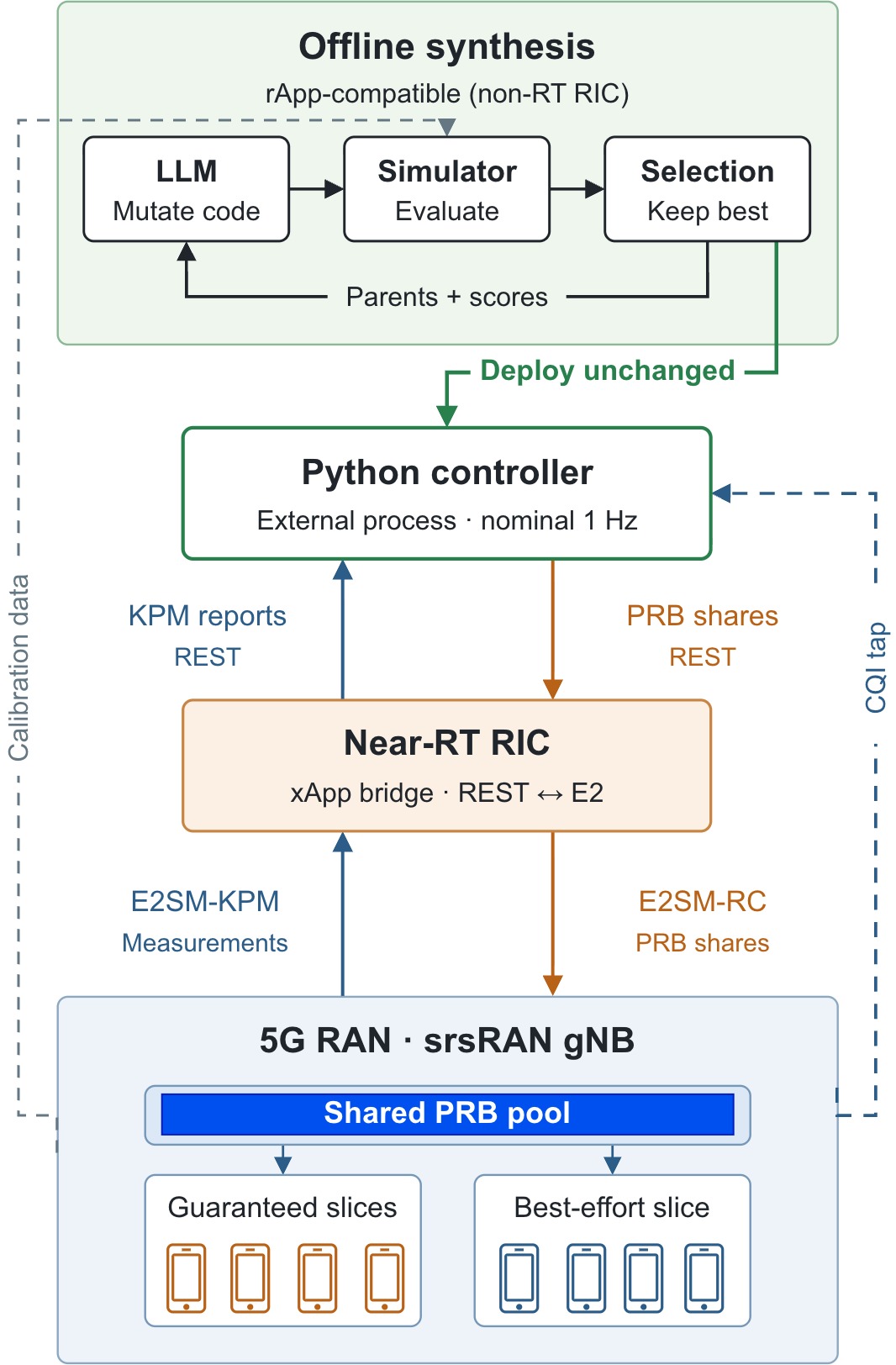}
\caption{Architecture of the proposed framework. Offline evolutionary synthesis produces a controller that executes in an external process connected to the near-RT RIC xApp.}
\label{fig:proposed_method}
\end{figure}

We apply LLM-guided evolutionary search to Python slicing controllers, following the program-synthesis paradigm demonstrated by FunSearch and AlphaEvolve for mathematical and computational problems~\cite{romera2024mathematical,novikov2025alphaevolve}. The LLM proposes programs and edits, while an external simulator measures fitness. Selection retains high-performing candidates, and the selected decision module executes unchanged in an external process connected to an xApp (Fig.~\ref{fig:proposed_method}). Applying it to O-RAN slicing raises domain-specific questions. Does the evolved controller actually transfer through a real E2 control path? Can an operator read its source and predict how it will respond to a channel fade? And does evolutionary search find better controllers than prompting the same LLM without fitness feedback?

The paper makes three empirical contributions:
\begin{enumerate}
\item \textbf{Evolved inspectable controllers for O-RAN slicing.} We introduce an offline synthesis framework that uses LLM-guided evolution to produce slicing controllers as compact Python programs whose allocation logic operators can read, trace, and edit after optimization. The evolved controller transfers to hardware unchanged, executing through the O-RAN control path on the POWDER 5G testbed.

\item \textbf{Inspectable control logic and repair.} We demonstrate that the inspectability of the evolved controllers makes decision rules traceable to runtime behavior, allowing operators to diagnose structural defects and apply targeted edits without retraining or re-evolving.

\item \textbf{Search performance vs. direct prompting.} We show that evolutionary search achieves higher average evaluation scores than independent prompting at a matched proposal budget, and that seeding evolution with a hand-tuned expert rule produces controllers that exceed the rule's own fitness.
\end{enumerate}

We validate the framework in two settings: a two-slice experiment on the NSF POWDER 5G testbed and a four-slice simulation using generated channel traces and a delivery model anchored to the same testbed calibration.
The simulation code, evaluation scripts, evolved
controller source, and testbed data are publicly
available on \href{https://github.com/ANRGUSC/evolving-oran-slicing-xapps}{https://github.com/ANRGUSC/evolving-oran-slicing-xapps}.

The remainder of this paper is organized as follows: Section~\ref{sec:related} frames our contributions against prior work; Section~\ref{sec:problem} defines the slice model and objective; Section~\ref{sec:framework} describes the synthesis procedure and its integration into the control path; Sections~\ref{sec:hardware} and~\ref{sec:evolution} present the hardware and simulation results; and Section~\ref{sec:conclusion} concludes.
\section{Related Work}
\label{sec:related}
\textbf{Experimental O-RAN control.} NexRAN demonstrates closed-loop slicing on POWDER with a software RAN, a RIC, and a custom slicing service model~\cite{johnson2022nexran}. OpenRAN Gym provides a workflow for data collection, xApp development, and experimental evaluation~\cite{bonati2023openran}. ColO-RAN develops learned control on programmable experimental infrastructure~\cite{polese2022colo}. These systems establish the platform capabilities on
which our work is based; our contribution is an
application-level study of evolved decision programs
and their failure modes.

\textbf{Learning and explanations for slicing.}
Several works apply DRL to O-RAN slice allocation:
DORA uses proximal policy optimization
(PPO)~\cite{dorcheh2025dora}, xSlice adds graph
representations and evaluates on an O-RAN
testbed~\cite{yan2025near}, and PandORA compares reward
designs and control timescales~\cite{tsampazi2024pandora}. EXPLORA
adds post-hoc explanations by producing human-readable
summaries of DRL actions and using them for action
steering~\cite{fiandrino2023explora}. In these approaches, the policy remains a neural
network whose allocation rules are encoded in learned
parameters. Our approach expresses the allocation
logic as readable source code that operators can
inspect and edit directly.

\textbf{LLMs for networking.} LLM-xApp places an LLM
directly in the control loop for radio resource
management~\cite{wu2025llm}. AutORAN uses an LLM to
generate xApps from natural-language specifications
and validates them on a real O-RAN
testbed~\cite{li2026autoran}. Our use of LLMs differs in
that the model is involved only during offline search,
avoiding runtime latency and dependence on external
inference.

\textbf{Executable program search.} FunSearch combines language-model proposals with executable evaluation and selection~\cite{romera2024mathematical}; AlphaEvolve extends the paradigm to scientific and computational optimization~\cite{novikov2025alphaevolve}. We apply this mechanism to O-RAN slicing. The networking-specific questions are whether the resulting program transfers through a real control path, whether source inspection explains its response to scarcity, and what allocation quality the measured search procedure obtains.

\section{System Model and Problem Formulation}
\label{sec:problem}

\subsection{Slice Allocation and Utility Objective}
\label{subsec:objective}

We consider an O-RAN downlink cell serving $m$ guaranteed slices $\mathcal P=\{1,\ldots,m\}$ and one residual best-effort (BE) slice $b$. At control tick $t$ the controller assigns integer physical resource block (PRB) percentage shares $s_i(t)$ to each guaranteed slice from a 100-point pool, and best-effort traffic receives the remainder:
\begin{equation}
s_b(t)=100-\sum_{i\in\mathcal P}s_i(t).
\label{eq:residual}
\end{equation}
Shares are integers bounded per slice and in aggregate,
\begin{equation}
\begin{gathered}
s_i(t)\in\mathbb Z_{\ge0},\qquad s_i^{\min}\le s_i(t)\le s_i^{\max},\\
\sum_{i\in\mathcal P}s_i(t)\le s^{\max}\le 100,
\end{gathered}
\label{eq:action}
\end{equation}
so that at least $100-s^{\max}$ points of the pool always remain with best effort. Releasing a guaranteed slice means clamping it to its minimum admissible share $s_i^{\min}$, which returns the rest of its allocation to best effort through (\ref{eq:residual}). The bounds are instance-specific and are stated with each evaluation configuration in Sections~\ref{sec:hardware} and~\ref{sec:evolution}.

The score rewards SLA attainment and best-effort throughput while penalizing allocation changes. A candidate policy $\pi$ is scored over an episode of $T$ ticks by
\begin{equation}
\mathcal C(\pi)=\frac1T\sum_{t=0}^{T-1}\Big[\sum_{i\in\mathcal P}w_iA_i(t)+\lambda_b\frac{r_b(t)}{R_b}-\lambda_c\,\chi(t)\Big],
\label{eq:objective}
\end{equation}
where $w_i$ is the slice weight, $A_i(t)\in\{1,-\tfrac12\}$ is binary demand-capped attainment, $r_b(t)$ is delivered best-effort throughput, $R_b$ a normalization ceiling, $\chi(t)$ the number of changed commands, $\lambda_b=0.6$, and $\lambda_c=0.02$. Attainment uses a $0.5$~Mbps tolerance: $A_i(t)=1$ if $r_i(t)\ge\min\{d_i(t),\tau_i\}-0.5$ for target $\tau_i$ and offered demand $d_i(t)$, and $-\tfrac12$ otherwise. Meeting the threshold increases the slice's attainment term by $1.5\,w_i$ relative to missing it. The ceiling $R_b$, the demand profile, and the set of commands counted by $\chi$ are instance parameters; scores are compared only within a configuration.

\textit{Operating regimes.} Let $a_i(t)=\max\{0,\min[d_i(t),\tau_i]-0.5\}$ be the demand-capped attainment threshold. Under a linear delivery model with per-share efficiency $e_i(t)>0$ (Mbps per percentage point), the minimum share that attains it, which we call the slice's \emph{price}, is
\begin{equation}
p_i(t)=\max\Big\{s_i^{\min},\Big\lceil\frac{a_i(t)}{e_i(t)}\Big\rceil\Big\};
\label{eq:price}
\end{equation}
for the affine deep-null model of (\ref{eq:delivery}) in Section~\ref{subsec:calib}, the price is read from the delivery function directly.
The controller must distinguish three operating regimes using lagged measurements:
\begin{enumerate}
\item \textbf{Abundant capacity:} $p_i(t)\le s_i^{\max}$ for every $i$ and $\sum_{i\in\mathcal P}p_i(t)\le s^{\max}$. Every target is attainable; the objective rewards holding each slice at its minimum viable share and returning the rest to best effort.
\item \textbf{Contention across guarantees:} $p_i(t)\le s_i^{\max}$ for every $i$ but $\sum_{i\in\mathcal P}p_i(t)>s^{\max}$. The objective rewards funding the subset of targets that maximizes weighted attainment plus residual credit, which in general is not the weight-ordered subset.
\item \textbf{Individual infeasibility:}
$p_k(t)>s_k^{\max}$ for some slice $k$.
Its attainment credit is $-\tfrac12$ at every admissible
share. Releasing it can increase best-effort throughput;
when its channel recovers, the controller can reconsider
funding its target.
\end{enumerate}

The preferred response depends on the utility, the allocation constraints, and the switching cost; release is rewarded under the binary credit of (\ref{eq:objective}), and an SLA that pays for partial delivery would change the preferred allocation and the scoring function accordingly.

To express improvement relative to a baseline and a reference within each evaluation setting, we define normalized gain as
\begin{equation}
\eta(\pi)=\frac{F(\pi)-F_{\rm base}}{F_{\rm ref}-F_{\rm base}},
\label{eq:eta}
\end{equation}
Here, $F(\pi)$ is the evaluation score used for the
comparison, and $F_{\rm base}$ and $F_{\rm ref}$ are
the corresponding scores of the baseline and reference
policies. All three scores use the same traces and
aggregation rule. Thus $\eta=0$ at the baseline and
$\eta=1$ at the reference; values can lie outside this
interval. The reference has access to the true current
channel state. The evaluation score and anchors are
specified with each comparison.

The two-slice hardware experiment
(Section~\ref{sec:hardware}) examines abundant capacity
under good channel conditions and individual infeasibility
during deep fades. The four-slice simulation
(Section~\ref{sec:evolution}) adds contention among
guaranteed slices. Their channels fade at different times,
and the resources needed to meet all targets exceed the
shared pool. The controller must therefore choose which
targets to satisfy.

\section{LLM-Guided Synthesis Framework}
\label{sec:framework}

\subsection{Evolutionary Search Procedure}
\label{subsec:search}

\begin{figure}[t]
\centering
\includegraphics[width=0.75\columnwidth]{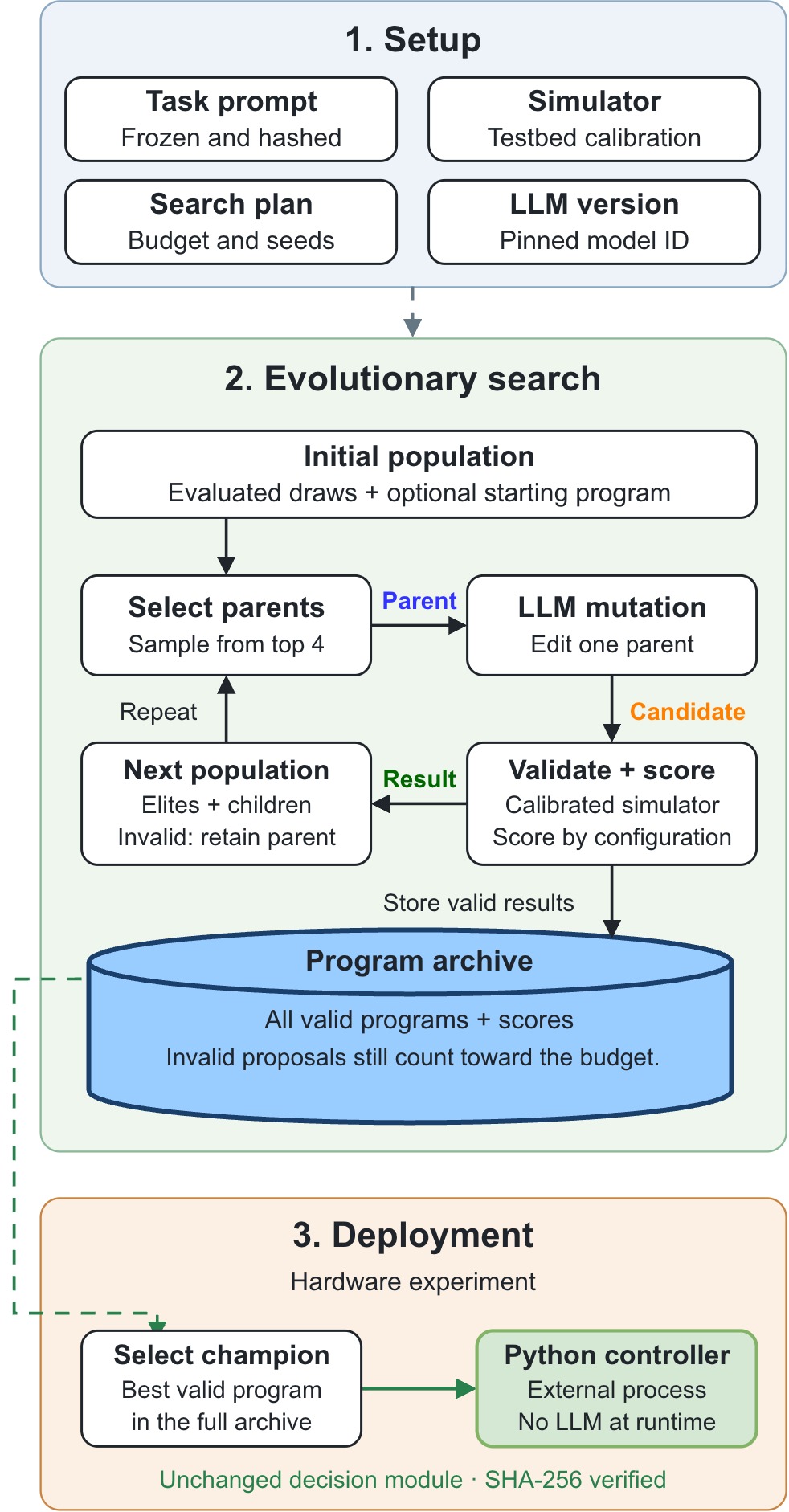}
\caption{Synthesis pipeline. Before search, the task prompt, model version, and evaluation protocol are fixed, and the simulator is calibrated against hardware. The LLM revises elite programs from the current population using their source, fitness, and summarized execution diagnostics. Candidates are evaluated on a fixed set of simulation traces; valid programs enter the archive, and invalid proposals are logged. The highest-fitness valid program in the archive is selected for final evaluation. In the hardware experiment, the selected decision module is deployed unchanged in the external control process, with source identity verified by hash. The held-out split was sealed throughout the primary campaign.}

\label{fig:pipeline}
\end{figure}

A \emph{proposal} is one candidate program generated
by the LLM, either from the task prompt or by revising
a parent. Invalid proposals still count toward the
budget. A \emph{search run} is one complete execution
of the search procedure.

Each candidate implements
\begin{equation}
s(t)=\text{controller}(\text{history},\text{state},\text{sla},\text{predictors}),
\label{eq:interface}
\end{equation}

and is scored by execution in the calibrated simulator.
The \texttt{predictors} argument is empty in all reported
experiments but could accommodate learned throughput
models, such as GNN predictors for multi-hop
$p$-CSMA networks~\cite{11568222}.
Fig.~\ref{fig:pipeline} shows the pipeline, and
Algorithm~\ref{alg:evolution} gives one evolution run.

Three methods are compared at the same proposal budget. \emph{Prompting alone} draws $B$ programs from the task prompt with no parent programs and no fitness feedback and keeps the best. \emph{Evolution from scratch} draws $n_0$ initial programs and then performs $G$ rounds of $k$ mutation proposals, giving $B=n_0+Gk$. \emph{Evolution from a starting program} uses the same proposal budget and additionally places a fixed starting program in the initial pool and archive. The starting program is supplied through the population; its source is included in a mutation prompt when it is selected as the parent.  In each round the top four programs are retained as elites and mutation parents are sampled uniformly from them. Mutation prompts contain the task specification, the parent's source, its fitness, and summarized execution diagnostics (per-slice attainment, violation rate, mean best-effort throughput, allocation changes). The task prompt specifies the interface, objective, and environment facts without prescribing an allocation rule.  Final selection takes the best valid record in the complete archive, so a seeded run cannot return a program with lower fitness on the search traces than its starting program. The four-slice experiments use five starting programs, including the fixed-threshold baseline defined in Section~\ref{subsec:baselines}.

\textit{Evaluation safeguards.} Before any draw, the prompt is sealed and hashed, the simulator is calibrated against the testbed (Section~\ref{subsec:calib}), the runs and their success criteria are pre-declared, and all four-slice searches use the fixed model version \texttt{claude-haiku-4-5-20251001}, so that what a controller was told can be separated from what the search discovered. In the four-slice experiment, we evaluate every candidate on the same 16 simulated channel traces, generated using random seeds 0–15. Each trace lasts 360 control ticks. We average the score over each trace and use the lowest of the 16 averages as the candidate’s fitness.
Every eligible candidate is an atomic record holding its source, finite fitness, validity, and provenance. In the four-slice evaluator, failed loads, missing
controller entry points, timeouts, and malformed or
invalid action responses invalidate the candidate.
Requests that pass validation are rounded to
nonnegative integer shares. If their sum exceeds the
guaranteed cap, they are scaled proportionally, with
remaining integer points assigned by largest remainder.
An invalid mutation still consumes its proposal
allowance, but its population slot retains the complete
parent record, including the parent's source and score. Each fault detector is validated by injecting a known fault. A trusted parent process owns the environment, random state, reward computation, and accounting; a sandboxed worker executes the candidate and returns only actions, and every response is journaled before evaluation. We use these same traces to select controllers and report the four-slice simulation results. A separate set of traces was kept sealed during all searches and used only for the post-hoc evaluation reported in Table~\ref{tab:matched}.

\begin{algorithm}[t]
\caption{One evolution run}
\label{alg:evolution}
\begin{algorithmic}[1]
\STATE Draw $n_0$ programs from sealed task prompt $P$; evaluate each and archive valid records.
\STATE If seeded, evaluate the fixed starting program and archive its record.
\STATE Initialize population $\Pi$ from the top eight valid initial records.
\FOR{round $g=1,\ldots,G$}
  \STATE Retain the top four records as elites $E$; set insertion list $I\leftarrow\emptyset$.
  \FOR{$j=1,\ldots,k$}
    \STATE Sample parent $p\sim\mathrm{Uniform}(E)$; request a mutation from its source and diagnostics.
    \STATE Journal the response; evaluate the child on the same simulation traces.
    \IF{the child is valid and its fitness is finite}
      \STATE Archive the child record; append it to $I$.
    \ELSE
      \STATE Log the failure; append the complete parent record to $I$.
    \ENDIF
  \ENDFOR
  \STATE $\Pi\leftarrow E\cup I$.
\ENDFOR
\STATE Return the highest-fitness valid record in the full archive.
\end{algorithmic}
\end{algorithm}

\subsection{Near-RT RIC Control-Path Integration}
\label{subsec:integration}

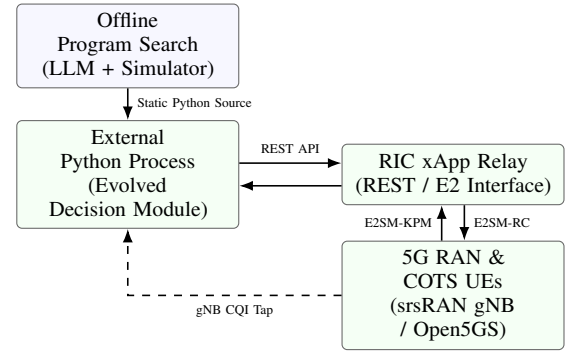
\begin{figure}[t]
\centering
\begin{tikzpicture}[
    font=\footnotesize,
    box/.style={draw=black!60, rounded corners=2pt, align=center, minimum height=8mm, text width=27mm, fill=blue!3},
    arr/.style={-{Latex[length=1.8mm]}, semithick}
]
\node[box] (search) at (0,0) {Offline Program Search\\(LLM + Simulator)};
\node[box, fill=green!4] (proc) at (0,-1.7) {External Python Process\\(Evolved Decision Module)};
\node[box, fill=green!4] (xapp) at (4.3,-1.7) {RIC xApp Relay\\(REST / E2 Interface)};
\node[box, fill=green!4] (ran) at (4.3,-3.3) {5G RAN \& COTS UEs\\(srsRAN gNB / Open5GS)};
\draw[arr] (search) -- node[right, font=\tiny] {Static Python Source} (proc);
\draw[arr] ([yshift=1.5mm]proc.east) -- node[above, font=\tiny] {REST API} ([yshift=1.5mm]xapp.west);
\draw[arr] ([yshift=-1.5mm]xapp.west) -- ([yshift=-1.5mm]proc.east);
\draw[arr] ([xshift=1.5mm]xapp.south) -- node[right, font=\tiny] {E2SM-RC} ([xshift=1.5mm]ran.north);
\draw[arr] ([xshift=-1.5mm]ran.north) -- node[left, font=\tiny] {E2SM-KPM} ([xshift=-1.5mm]xapp.south);
\draw[arr, dashed] (ran.west) -| node[pos=0.25, below, font=\tiny] {gNB CQI Tap} (proc.south);
\end{tikzpicture}
\caption{Implemented hardware control path. The evolved decision module executes in an external Python process connected to the near-RT RIC xApp through REST; the xApp relays commands and measurements over E2, and a gNB tap supplies CQI.}
\label{fig:architecture}
\end{figure}

Fig.~\ref{fig:architecture} shows the live execution path. The synthesized decision module is deployed unchanged into an external control process and verified at load by SHA-256 match against the offline champion. Each control period the process queries per-slice key performance measurements (KPMs) from the O-RAN Software Community (OSC) near-RT RIC xApp over REST, reads the live channel quality indicator (CQI) through a gNB tap, executes the deterministic decision logic, and posts PRB allocation commands back to the xApp, which packages them as E2 service model RAN control (E2SM-RC) messages to the srsRAN gNB scheduler~\cite{srsric}. The loop sleeps for one second after each decision, so one hertz is the nominal rather than the measured control rate. Unchanged deployment refers to this decision module; the adapter that supplies observations and delivers commands is platform-specific. The synthesis loop is a non-real-time function compatible with an rApp on the non-RT RIC; this paper validates one iteration of the cycle (calibrate, evolve, predict, deploy, confirm), and periodic re-synthesis is future work.
\section{Hardware Transfer and Inspectable Decisions}
\label{sec:hardware}

We first describe the testbed and calibrated simulator, summarize the two-slice searches, and introduce the PPO comparator. We then inspect the controllers' source code to predict release behavior and identify a state-update defect. We check these predictions on a fixed CQI sequence and compare the predicted release and non-release responses with the hardware observations. Finally, we test a one-line calibration repair in a separate simulation.

\subsection{Hardware Testbed Setup}
\label{subsec:testbed}

\begin{figure}[t]
\centering
\includegraphics[width=0.73\columnwidth]{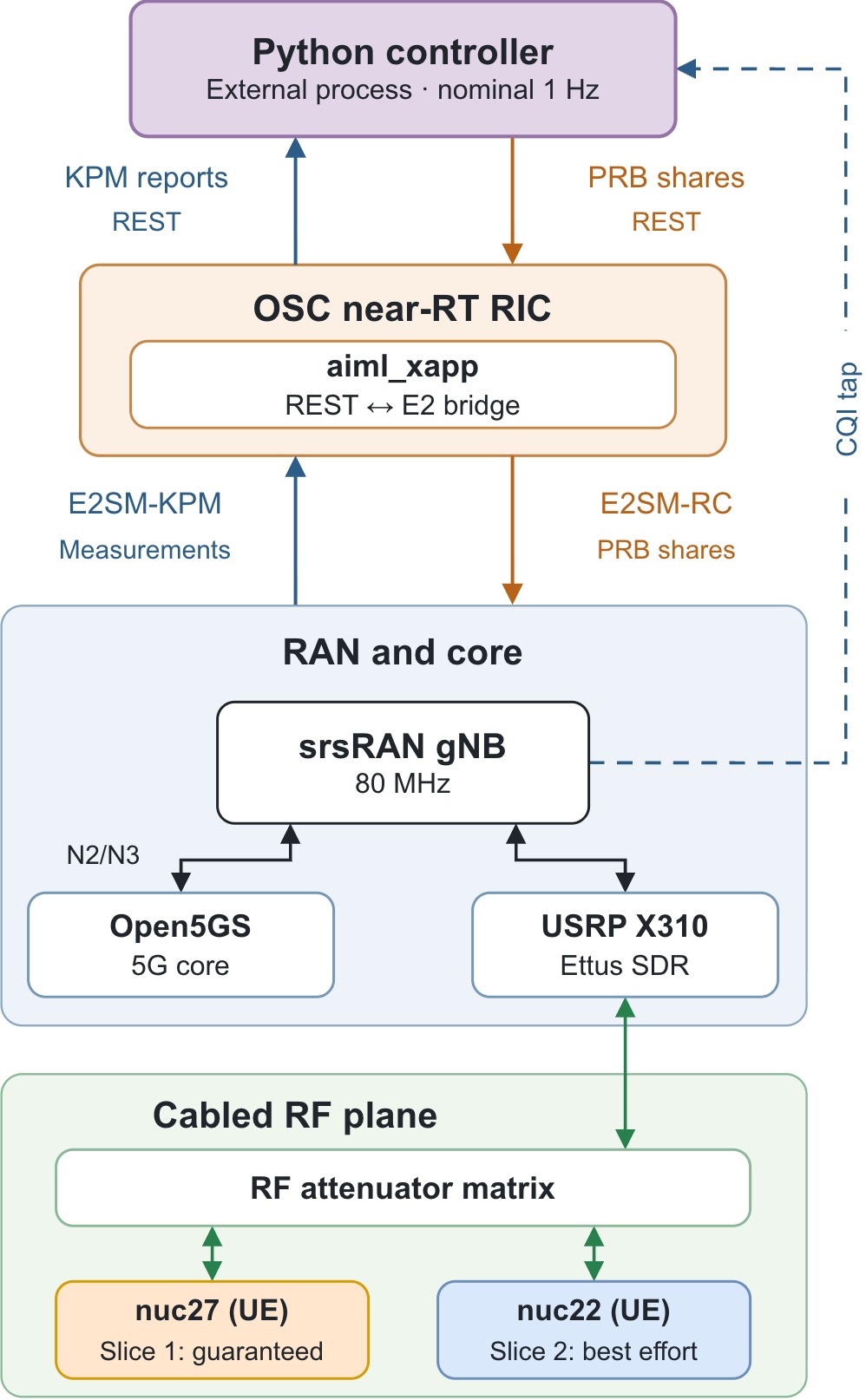}
\caption{POWDER testbed architecture. The RF plane is cabled through a programmable attenuator matrix to two commercial off-the-shelf (COTS) UEs, one per slice; deep nulls are induced by attenuating the guaranteed-slice UE's path. The software control path is detailed in Fig.~\ref{fig:architecture}.}
\label{fig:testbed_architecture}
\end{figure}

Experiments are conducted on the NSF POWDER testbed~\cite{powder} over cabled RF links. As illustrated in Fig.~\ref{fig:testbed_architecture}, the hardware platform comprises:
\begin{itemize}
    \item \textbf{RAN platform:} an srsRAN 5G gNB operating at $80$~MHz bandwidth on an Ettus Research USRP X310 software-defined radio (SDR), connected to an Open5GS 5G core over the N2/N3 interfaces.
    \item \textbf{Control plane:} the OSC near-RT RIC hosting an xApp REST bridge (\texttt{aiml\_xapp}) that relays E2SM-KPM reports and E2SM-RC control commands; the evolved decision module runs in an external process that queries the bridge each control period.
    \item \textbf{RF environment:} two commercial 5G user equipment (UE) devices cabled through a programmable RF attenuator matrix. \texttt{nuc27} serves Slice~1 (guaranteed, $80$~Mbps target) and \texttt{nuc22} serves Slice~2 (best effort). The guaranteed share ranges over $s_1\in\{5,\ldots,90\}$.
\end{itemize}

\textit{Measurement protocol.} Each controller in
Table~\ref{tab:panel_responses} was evaluated once under
each channel condition during the same hardware session.
Reports were sampled approximately once per second,
with a mean inter-sample interval of 1.07~s.
Each run lasted approximately 120~s and contained about
110 samples. Good-channel and deep-null runs used fixed
attenuator settings, with CQI approximately 5 after the
null was established. Reported PRB shares and throughput
values are arithmetic means over all logged samples
in each run, including startup transients.

In the static 90/10 deep-null run, the guaranteed
slice's mean throughput is 37.1~Mbps at a 90\%
commanded PRB share, well below its 80~Mbps target.
Performance is evaluated against the static 35/65
allocation, the best static split identified for this
setup in the calibrated simulator, which scores
$0.9300$. This comparison measures the benefit of
adapting allocations when channel conditions change.

\subsection{POMDP Formulation and Testbed Calibration}
\label{subsec:calib}

The controller receives channel-quality reports and past throughput measurements, while true efficiency $e_i(t)$ is not observable over E2. We therefore model slicing control as a partially observed Markov decision process (POMDP). At tick $t$ the controller observes
\begin{equation}
o(t)=\{q_i(t),\,r_i(t-1),\,s_i(t-1)\}_{i\in\mathcal P\cup\{b\}},
\label{eq:obs}
\end{equation}
where $q_i(t)$ is the reported wideband CQI and $r_i(t-1)$ the previous delivered throughput. The policy is a map $(s(t),x_{t+1})=\pi(o(t),x_t)$ from the observation and persistent program state $x_t$ to the share vector; the state lets a program carry filtered measurements, recent observations, or a previously selected subset across ticks. The evaluator has access to true current demand and channel state; the controller uses the observations in (\ref{eq:obs}).

The two-slice simulator, which is calibrated to the testbed, follows a predefined schedule of channel conditions. Delivery to the guaranteed slice is:

\begin{equation}
r_1(s)=\begin{cases}\min\{2.39\,s\,(1+\epsilon),\,d_1\}, & \text{good regime},\\ \max\{0,\,0.567\,s-8.1\}\,(1+\epsilon), & \text{deep-null regime},\end{cases}
\label{eq:delivery}
\end{equation}
with multiplicative noise $\epsilon\sim\mathcal N(0,0.0302^2)$, and best-effort delivery is $r_b=2.107(100-s_1)+31.2$~Mbps. Both curves are fitted to POWDER measurements collected under the protocol of Section~\ref{subsec:testbed}; they are calibrated approximations, and measured radio goodput is neither deterministic nor exactly linear in share. CQI observations are noisy proxies of the regime: good-regime samples are clipped Gaussian around 15 ($\sigma=0.79$), and deep-null samples are centered on a floor of 4.5, 5.1, or 5.8 ($\sigma=0.5$), clipped to $[1,15]$. The controller's action takes effect at the current tick. A fitness evaluation averages over two best-effort demand profiles and 16 seeds for each floor and takes the minimum over the three floors, so it uses 96 episodes and rewards controllers that survive the worst floor rather than the average one. The four-slice scenario replaces the two-regime schedule with per-slice fading through a constructed efficiency curve anchored to the testbed calibration, described in Section~\ref{sec:evolution}.

\subsection{Search Statistics}
\label{subsec:search-stats}

To assess sensitivity to the generating model, we also ran the two-slice search with Gemini (\texttt{gemini-2.5-flash-lite}) as the proposal LLM. The two model families differed substantially in search quality. Gemini voided $65\%$ of its generated programs on load or execution faults. Claude voided $5\%$ and improved by $0.034$ over the generations.

Prompt wording also affected initial proposals. A confident statement of the environment facts led $13$ of $20$ initial programs to propose a release rule, compared with $3$ of $20$ under a hedged framing. However, no prompt wording produced a correctly placed threshold. Evolutionary search selected a controller with an effective release threshold.


\subsection{PPO Comparator}
\label{subsec:ppo}

As a learned baseline, we train a proximal policy
optimization (PPO) agent on the same calibrated two-slice
simulator. The agent uses Stable-Baselines3 with an MLP
policy network, the observation of~(\ref{eq:obs}), the
integer-share action of~(\ref{eq:action}), and the
composite reward of~(\ref{eq:objective}). It therefore
optimizes the same objective as the evolved controllers
but exposes its policy only as network weights.

We train five PPO agents using different random seeds for $10^6$ environment
steps each. All five learn the release behavior (settled
guaranteed share of 5.0\% during the deep null). The
fitness (minimum over the three CQI floors) is $0.9534 \pm 0.0208$ across seeds,
exceeding the static baseline score (0.9300) in four of five. We
deploy the best-scoring trained model through the same
hardware control path and measurement protocol as the
evolved controller.

\subsection{Selected Program and Source-Based Predictions}
\label{subsec:predictions}

Each candidate is evaluated at three CQI floors: 4.5, 5.1, and 5.8. At each floor, scores are averaged over two best-effort traffic profiles and 16 random seeds; the evaluation score is the lowest of the three averages. None of 144 zero-shot programs from two model families outperforms the static baseline of $0.9300$; the best scores $0.9054$. Evolutionary search produced a controller scoring $0.9398$. That program is the module analyzed and deployed below.

The evolved controller consists of 98 lines of Python. Its decision logic evaluates a two-stage release predicate:
\begin{align}
\bar c_t &= 0.3\,\widetilde c_t + 0.7\,\bar c_{t-1}, \label{eq:ema} \\
z_t &= \mathbf{1}\{\bar c_t < 6.5\} \cdot \mathbf{1}\left\{\sum_{j=0}^{2} \mathbf{1}\{\widetilde c_{t-j} < 6\} \ge 2\right\}, \label{eq:trigger}
\end{align}
where $\widetilde c_t$ is the instantaneous wideband CQI and $\bar c_t$ is an exponential moving average (EMA) initialized on the first tick (Fig.~\ref{fig:excerpt}).

\begin{figure}[t]
\begin{lstlisting}
cqi_ema = 0.3 * current_cqi + 0.7 * cqi_ema
recent_low_cqi = sum(
    1 for c in cqi_history[-3:] if c < 6.0
) >= 2
is_deep_null = cqi_ema < 6.5 and recent_low_cqi
mbps_per_prb = max(0.08, mbps_per_prb)
prb_needed = target_mbps / mbps_per_prb
if is_deep_null:
    new_share = sla.priority.min_share
\end{lstlisting}
\caption{Executable decision logic extracted from the evolved controller. The explicit predicate isolates deep channel nulls and clamps the allocation to the minimum share.}
\label{fig:excerpt}
\end{figure}

When $z_t=1$, the program clamps the slice to its minimum share ($s_1^{\min}=5$). Outside deep nulls, the program estimates efficiency as $\max\{0.08,\,0.145\,\bar c_t+0.08\}$~Mbps per share point, maps the target rate to an integer share, clips to the SLA bounds, and updates the allocation only when the requested change exceeds a channel-dependent hysteresis of 1--3 points. The release predicate depends only on channel telemetry ($\widetilde c_t$) and ignores delivered throughput, so release cannot be distorted by throughput-estimation feedback; the same property means the provisioning rule trusts its calibrated efficiency estimate and cannot correct a delivery-model error from measurement.

From the source we derive three predictions and verify them on a deterministic 75-step CQI fixture (30 observations at CQI 15, 30 at CQI 5, 15 at CQI 15):
\begin{itemize}
    \item \textbf{Release timing:} from the CQI-15 steady state the EMA follows $\bar c=5+10(0.7)^k$ after $k$ low observations, so (\ref{eq:trigger}) first holds at $k=6$. Execution on the fixture releases on the sixth low observation, matching the prediction.
    \item \textbf{Immediate restoration:} with no latched release state and no separate exit threshold, the controller should restore a share above the floor on the first recovered observation. Execution confirms this.
    \item \textbf{Comparison controller defect:} inspection of the comparison controller (evolved with Gemini, \texttt{gemini-2.5-flash-lite}) reveals a state-update defect: a local variable overwrites the release state at the end of every tick, preventing the release path from ever executing. On the same fixture it never releases, even at a CQI floor of 2. This structural bug is invisible from the fitness score alone, which reports only low overall performance.
\end{itemize}

\subsection{Hardware Confirmation}
\label{subsec:hardware_confirm}

\begin{table}[t]
\centering
\small
\caption{Two-slice hardware results on the POWDER
testbed, with an 80~Mbps guaranteed-slice target.
Entries are full-run arithmetic means, including
startup samples. Share columns report the guaranteed
slice's commanded PRB share under each channel
condition; BE throughput is measured during the
deep-null run.}
\label{tab:panel_responses}
\resizebox{\columnwidth}{!}{%
\begin{tabular}{@{}lccc@{}}
\toprule
Controller &
\begin{tabular}[c]{@{}c@{}}
Guar.\ share\\ null (\%)
\end{tabular} &
\begin{tabular}[c]{@{}c@{}}
Guar.\ share\\ good ch.\ (\%)
\end{tabular} &
\begin{tabular}[c]{@{}c@{}}
BE tput\\ null (Mbps)
\end{tabular} \\
\midrule
Claude evolved  & 7.9  & 35.4 & 228.6 \\
RL policy (PPO) & 5.0  & 36.6 & 235.6 \\
Static (35/65)  & 35.0 & 35.0 & 158.2 \\
Gemini evolved  & 86.6 & 40.3 & 38.6  \\
Static (90/10)  & 90.0 & 90.0 & 30.1  \\
\bottomrule
\end{tabular}%
}
\end{table}

The selected controller executed unchanged on the POWDER
testbed. Its PRB allocation decisions passed from the
external Python process through the xApp bridge and
E2SM-RC to the gNB scheduler.

Table~\ref{tab:panel_responses} reports the hardware
results. The Claude-evolved controller releases
resources during the deep null and achieves a full-run
mean best-effort throughput of 228.6~Mbps, compared with
158.2~Mbps under the static 35/65 allocation, a 44.5\%
increase. The PPO policy trained on the same simulator
commands the minimum guaranteed share throughout the
null run and achieves 235.6~Mbps.

The comparison controller, evolved with Gemini, does
not release resources during the null run. Its mean
commanded guaranteed share is 86.6\%, compared with
40.3\% in its good-channel run, and its mean best-effort
throughput is 38.6~Mbps. This behavior is consistent
with the state-update defect identified from its source
in Section~\ref{subsec:predictions}. The static 90/10
allocation achieves 30.1~Mbps of best-effort throughput.
Both results are consistent with reduced best-effort
capacity when a large allocation is retained on the
faded guaranteed link.


\subsection{Editability: Repairing a Calibration Error}
\label{subsec:calibration-edit}

We tested whether a direct source edit could compensate
for a calibration mismatch in the Claude-evolved
controller. In an offline simulation, we reduced actual
good-channel efficiency by 10\% while leaving the
controller's efficiency estimate unchanged. Because the
controller still assumed the original efficiency, it
under-provisioned the guaranteed slice, raising SLA misses
from 2.4\% to 79.9\%. To compensate, we reduced the efficiency estimate used
for provisioning by 10\%, changing the denominator
from $\widehat e$ to $0.9\,\widehat e$. The
controller then allocates additional PRBs to the
guaranteed slice at the expense of best-effort capacity,
compensating for the efficiency shortfall. The edit
reduced SLA misses to 2.2\% with a corresponding
6.3~Mbps reduction in mean best-effort throughput, and
required no re-evolution.
\section{Evolving Coordination in Trace-Driven Simulation}
\label{sec:evolution}
This section moves from the two-slice hardware experiment to a four-slice scenario where guaranteed slices fade on different schedules, creating contention for a shared resource pool. The delivery model is anchored to the same POWDER calibration used in Section~\ref{sec:hardware}. Because the dynamics are known and the allocation space
is small, we can compute policies using the true current
channel as privileged references for comparison.

\subsection{Environment}
The scenario has four guaranteed slices with weights $(1.0,0.9,0.8,0.7)$ and throughput targets $(44,46,48,50)$~Mbps. They share a guaranteed cap of 70 share points, with at least 30 points always reserved for best effort. The scenario is deliberately over-subscribed: even under the best channel conditions (2.39~Mbps per share point), the minimum required shares are $p_i=\lceil(\tau_i-0.5)/2.39\rceil$, giving $(p_1,p_2,p_3,p_4)=(19,20,20,21)$. Their sum is 80, which exceeds the cap of 70. The four targets therefore cannot all be met at the same time. The controller must decide which targets to fund at each tick, using observations delayed by one tick.

Scoring uses $R_b=239$~Mbps and counts changes to all five commands. Each slice fades on its own 80-tick cycle (10 high-CQI ticks, a 30-tick decline, 10 low-CQI ticks, a 30-tick recovery), with the four slices offset by 20 ticks. Delivery is $r_i(t)=e(c_i(t))\,s_i(t)$ with the constructed CQI-to-efficiency curve below, whose good-channel endpoint of 2.39~Mbps per share point comes from the POWDER calibration:

\begin{equation}
e(c)=0.30+(2.39-0.30)\Big(\frac{\operatorname{clip}(c,5,15)-5}{10}\Big)^{1.578}.
\label{eq:eff}
\end{equation}

To introduce variation across channel traces, Gaussian noise ($\sigma=0.5$) is added to the CQI during the high and transitioning phases but not during the low-CQI floor, and all CQI values are clipped to $[5,15]$. The noise is seeded so that different controllers face the same channel conditions under the same seed. The controller observes the channel one tick late (except on the first tick, where it sees the initial state). Each evaluation runs 16 episodes (seeds 0--15) of 360 ticks each.

\subsection{References and Baselines}
\label{subsec:baselines}

Each policy is evaluated on 16 seeds of 360 ticks each.
Fitness is the worst episode score across seeds,
$F(\pi)=\min_{j\in\{0,\ldots,15\}}\mathcal C_j(\pi)$,
so a controller must handle the hardest trace, not just
the average.

The \emph{fixed-threshold baseline} computes allocations
independently per slice using the nominal efficiency
curve of~(\ref{eq:eff}) and fixed CQI thresholds
$(9,9,11,11)$, with no coordination across slices. It
scores $-0.0485$ and serves as one starting program for seeded evolution.

The \emph{oracle reference} has access to the true
current channel. Each tick it enumerates the 16 possible funded subsets,
discards those whose minimum required shares exceed
the cap, and selects the feasible subset with the
highest reward, scoring 1.0555. For
comparison, a weight-ordered policy that funds slices in priority
order with the same perfect channel knowledge scores
0.9104, which is 13.7\% lower. The evolved controllers do not
have this advantage; they observe the channel one tick
late.

The \emph{hand-written expert rule} uses the same lagged
observations as the evolved controllers. It knows the
nominal efficiency curve of~(\ref{eq:eff}), which is
absent from the task prompt but is encoded in the
fixed-threshold baseline starting program. Each tick, it evaluates all
feasible funded subsets and the option of retaining
the previous allocation, choosing the highest predicted
composite score including the change penalty. We tune
CQI extrapolation strength $\alpha\in\{0,0.5,1\}$ and
efficiency multiplier $m\in\{0.8,0.9,1.0\}$ on the
same 16 simulation traces used during search, giving nine settings. The best
configuration ($\alpha=0$, $m=0.9$) scores 0.541.

Normalized gain in~(\ref{eq:eta}) uses
$F_{\rm base}=-0.0485$ and $F_{\rm ref}=1.0555$, placing
the fixed-threshold baseline at $\eta=0\%$, the oracle reference
at $\eta=100\%$, and the expert rule at $\eta=53.4\%$.

\begin{table}[t]
\centering\footnotesize
\setlength{\tabcolsep}{3pt}
\caption{Anchor policies for the four-slice scenario.
Fitness is the worst episode score across 16 seeds of
360 ticks each. The oracle reference uses the true
current channel; the other two observe it one tick late.
$\eta$ in~(\ref{eq:eta}) places the fixed-threshold baseline
at $0\%$ and the oracle reference at $100\%$.}
\label{tab:refs}
\begin{tabular}{@{}lrr@{}}
\toprule
Policy & Fitness & $\eta$ (\%) \\
\midrule
Oracle reference (true current channel) & 1.0555 & 100.0 \\
Hand-written expert rule (lagged obs.)  & 0.541  & 53.4 \\
Fixed-threshold baseline (lagged obs.)         & $-0.0485$ & 0.0 \\
\bottomrule
\end{tabular}
\end{table}

\subsection{What Search Adds at a Matched Proposal Budget}
\label{subsec:matched}
\begin{figure}[t]
\centering
\includegraphics[width=\columnwidth]{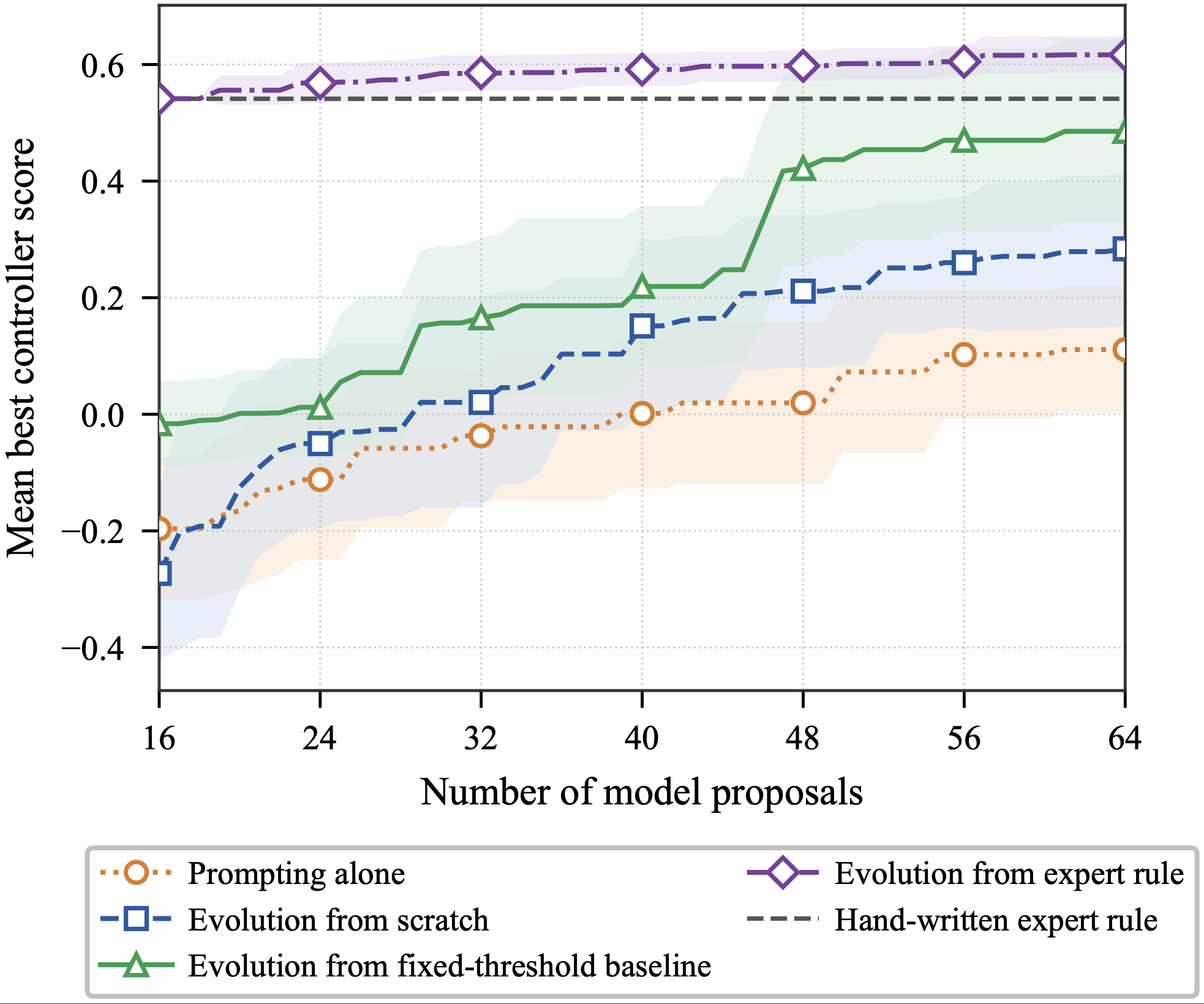}
\caption{Best fitness vs. proposal budget. All three evolutionary variants keep improving with more proposals; prompting alone gains little after the initial draws.}
\label{fig:convergence}
\end{figure}

To measure the contributions of fitness feedback and
starting programs, we evaluate prompting alone,
evolution from scratch, and evolution seeded with
five fixed programs. Each of these seven configurations
uses 64 LLM proposals per run and ten search runs,
with the same model, shared task prompt, evaluator,
and final selection rule. We use
\texttt{claude-haiku-4-5-20251001} at temperature 1.0.
This model keeps the cost of generating thousands of
candidate programs manageable. Holding the model fixed
allows us to compare the search procedures at the same
model scale. Evolution draws $n_0=16$ initial programs,
then runs $G=12$ rounds of $k=4$ mutation proposals.
Each run returns the valid program with the highest
fitness on the 16 search traces. For seeded evolution,
the proposal budget measures the additional search
after supplying the starting program.

The primary comparison covers prompting alone and
evolution from scratch. The starting-program experiments
then seed evolution with five programs of increasing
fitness, from the fixed-threshold baseline ($-0.048$) to
a previously evolved champion ($0.678$), to test how the
starting point affects the outcome.
Fig.~\ref{fig:convergence} shows that all methods
improve their best fitness as more candidates are
proposed; the curves track the best program found so
far, rather than the quality of each individual
proposal. Table~\ref{tab:matched} reports, for each
method, the returned fitness (minimum over the 16 search
traces) averaged across runs, and the mean and standard
deviation of held-out performance.

\begin{table}[t]
\centering
\scriptsize
\setlength{\tabcolsep}{3pt}
\caption{Four-slice results (64 proposals/run; 10 search
runs/method). Selection is the minimum episode reward
over 16 search traces; held-out is the mean over 200
unseen traces with new CQI noise. Parentheses give starting seed selection fitness.
$\eta_{\mathrm{held}}$: $0\%$ at the fixed-threshold
baseline and $100\%$ at the reference policy.}
\label{tab:matched}

\begin{tabularx}{\columnwidth}{
@{}>{\raggedright\arraybackslash}Xrrr@{}}
\toprule
Method (starting fitness)
& \shortstack{Selection\\min.}
& \shortstack{Held-out\\mean}
& $\eta_{\mathrm{held}}$ (\%) \\
\midrule

Prompting alone
& $0.111$
& $0.158\pm0.160$
& 16.3 \\

Evolution from scratch
& $0.284$
& $0.327\pm0.192$
& 32.1 \\
\midrule
Evolution: fixed-threshold seed ($-0.048$)
& $0.485$
& $0.529\pm0.219$
& 51.0 \\

Evolution: greedy seed ($0.204$)
& $0.564$
& $0.608\pm0.081$
& 58.5 \\

Evolution: prompting seed ($0.297$)
& $0.394$
& $0.436\pm0.104$
& 42.3 \\

Evolution: expert rule ($0.541$)
& $0.617$
& $0.677\pm0.045$
& 64.9 \\

Evolution: champion seed ($0.678$)
& $0.681$
& $0.712\pm0.007$
& 68.2 \\

\midrule

Hand-written expert rule
& $0.541$
& $0.613$
& 58.9 \\

\bottomrule
\end{tabularx}
\end{table}


On the held-out traces, evolution from scratch achieves
a higher mean than prompting alone (0.327 vs.\ 0.158).
Both remain below the hand-written expert rule (0.613).
Seeding closes this gap: evolution from the expert rule
reaches 0.677 and from the champion 0.712, both above
the expert rule itself.
We compute $\eta_{\mathrm{held}}$ from mean held-out reward using anchors $-0.0164$
for the fixed-threshold baseline and $1.0523$ for
the reference policy.
 Section~\ref{subsec:seed-choice}
examines how the choice of starting program drives
these outcomes.

\subsection{How the Starting Program Affects Search}
\label{subsec:seed-choice}

Two findings emerge from the starting-program
experiments in Table~\ref{tab:matched}.

First, evolution can improve on a strong starting point.
Seeding with the expert rule (0.541) yields a mean of
0.617, an average gain of 0.076 over the starting
program. Seeding with the champion (0.678) yields 0.681.

Second, a simpler starting program can evolve further than
a higher-scoring but more complex one. The compact greedy
seed starts at 0.204 and evolves to 0.564. The prompting
seed starts higher at 0.297 but evolves to only 0.394.
The two seeds differ in length (40 vs.\ 134 lines),
structure, and the knowledge they encode, so this
comparison does not isolate which factor drives the
difference.

\subsection{Understanding Controller Behavior}
\label{subsec:mechanism}

\begin{figure}[t]
\centering
\includegraphics[width=\columnwidth]{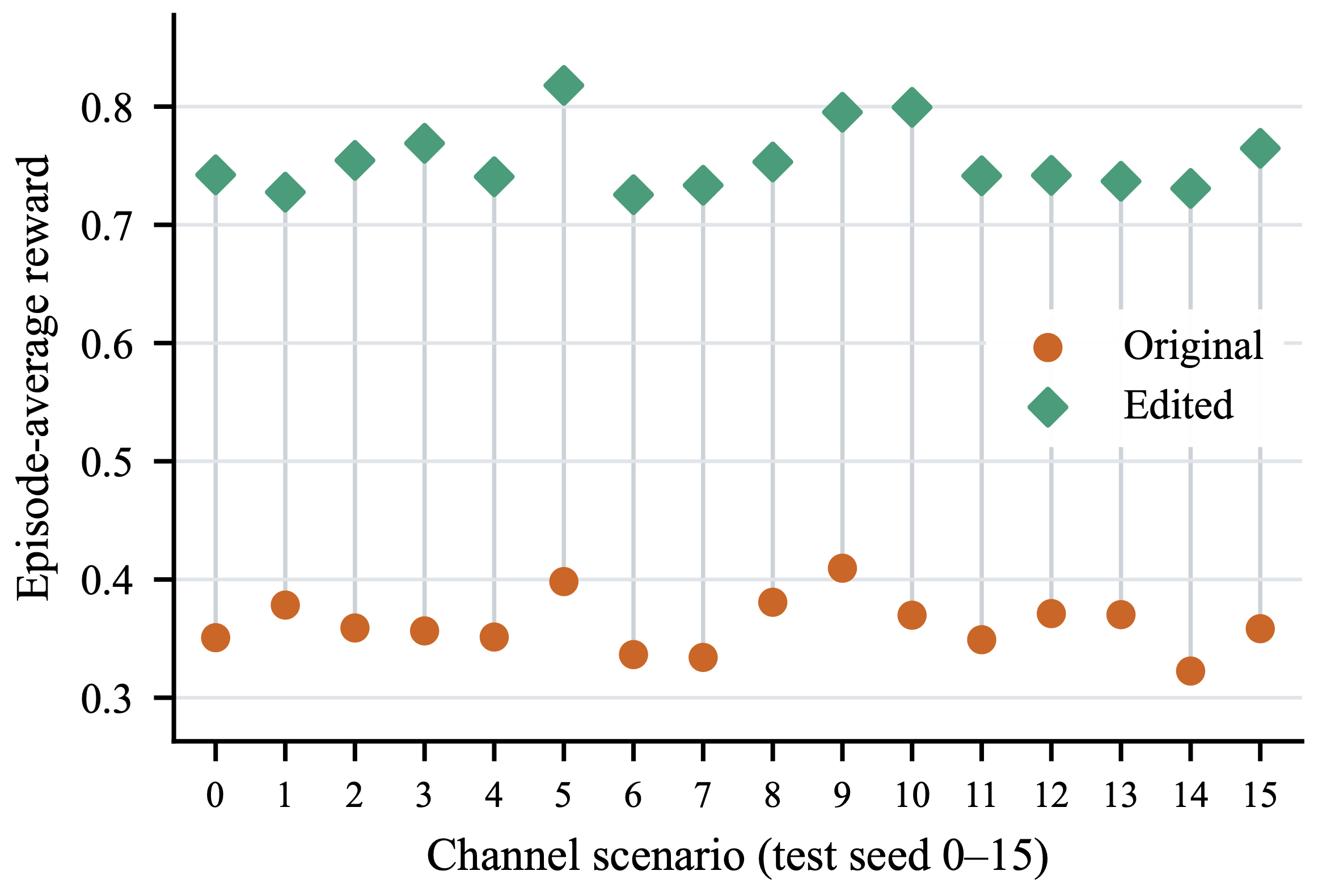}
\caption{One parameter source repair across 16 channel scenarios.
Each point is the episode reward (mean per-tick reward
over 360 ticks) before and after correcting a single
rate coefficient (11.8~$\to$~2.19) in a saved evolved
controller. Every trace improves.}
\label{fig:source-repair}
\end{figure}

Evolution does not always produce correct internal
estimates. We inspected five controllers from the
search comparison whose fitness fell below the expert
rule and found that each contained an error that is numerically identifiable by a human in its rate estimation or allocation
logic. We illustrate one of these repairs in
Fig.~\ref{fig:source-repair}, which compares the
controller before and after correction on all
16 evaluation traces.

This controller, returned by the prompting-seeded
evolutionary arm, estimates per-slice resource costs
to prioritize allocations. When a slice has little or
no allocation, it falls back to a CQI-based rate
estimate:
\begin{equation}
\widehat e(c)=\max\!\left(0.2,\;0.2+a\,\frac{c-5}{10}\right).
\label{eq:fallback}
\end{equation}
Reading the source reveals the problem: the coefficient
$a=11.8$ predicts 12~Mbps per share point at CQI~15,
five times the simulator's actual maximum of 2.39. The
controller therefore underestimates how many PRBs it
needs in good channels. Replacing 11.8 with 2.19 fixes
the high-CQI endpoint without changing anything else
in the controller.

We ran both versions on the same 16 simulation traces used during search.
Worst-episode fitness rose from 0.3226 to 0.7255, and
every trace improved. Target attainment increased from
37.7\% to 45.2\%, allocation churn fell from 3.76 to
2.26 changes per tick, and restoration improved from
41.6\% to 50.5\%. Best-effort throughput decreased by 3~Mbps because more PRBs were allocated to guaranteed slices.

We made this edit after the search and evaluated it on the same simulation traces. We report the repair separately from the search comparison. No further evolutionary search or retraining was needed.

\section{Conclusion}
\label{sec:conclusion}

LLM-guided evolution produced compact Python slicing
controllers whose source code operators can read, edit,
and deploy. On the NSF POWDER 5G testbed, the evolved
controller ran unchanged through the O-RAN control path
and released resources during a sustained channel fade.
Its mean best-effort throughput was 44.5\% higher than
the static comparator and close to that of the PPO
policy trained on the same simulator (within 3\%).

Source inspection supported both prediction and repair.
Reading the evolved controller's code predicted its
release behavior, confirmed in simulation and on
hardware. Inspecting a comparison controller (Gemini)
exposed a state-update defect that fitness scores alone
could not localize. A one-line calibration edit reduced
SLA misses from 79.9\% to 2.2\% without re-evolving.

In the four-slice simulation, evolutionary search
achieved higher average evaluation scores than
independent prompting at the same proposal budget, and
seeding evolution with the expert rule produced
controllers that exceeded the expert's own fitness.
Separately, a one-parameter source repair on a saved
controller corrected a five-fold rate overestimate and
more than doubled its fitness.

Future work will extend hardware validation to more UEs
and a four-slice configuration, and run the synthesis
loop periodically as a non-RT RIC function with
calibration drift as its trigger. The simulation code,
evaluation scripts, evolved controllers, and testbed
data are available in our
\href{https://github.com/ANRGUSC/evolving-oran-slicing-xapps}{public repository}.

\section*{Acknowledgments}

We thank the POWDER team and technical staff for testbed
access, training, and research support through the NSF
CyberPowder Fellows Program. This work used the POWDER
platform. The first author was supported by NSF CyberTraining
award 2417934.

\section*{AI Use Statement}
Claude (Anthropic) was used as the proposal model in
the evolutionary search pipeline described in
Section~\ref{sec:framework}. Gemini (Google) was used
as a comparison model in the two-slice search
(Section~\ref{subsec:search-stats}). Claude was also
used to assist with manuscript editing. The human
authors accept full responsibility for the contents
of this paper.

\bibliographystyle{IEEEtran}
\bibliography{references}
\appendices
\lstdefinestyle{controllerAppendixPrompt}{
  language={},basicstyle=\ttfamily\footnotesize,
  columns=fullflexible,keepspaces=true,showstringspaces=false,
  breaklines=true,breakatwhitespace=false,
  frame=single,framerule=0.3pt,rulecolor=\color{black!45},
  xleftmargin=4pt,xrightmargin=4pt,numbers=none,
  aboveskip=3pt,belowskip=8pt,
  literate={–}{{-}}1 {—}{{--}}2
}
\lstdefinestyle{controllerAppendixCode}{
  language=Python,basicstyle=\ttfamily\footnotesize,
  columns=fullflexible,keepspaces=true,showstringspaces=false,
  breaklines=true,breakatwhitespace=false,
  frame=single,framerule=0.3pt,rulecolor=\color{black!45},
  numbers=left,numberstyle=\tiny\color{black!60},numbersep=6pt,
  xleftmargin=18pt,xrightmargin=4pt,
  commentstyle=\color{black!65},keywordstyle=\bfseries,
  stringstyle=\color{black},aboveskip=3pt,belowskip=8pt
}
\newcounter{controllerAppendixListing}
\newcommand{\controllerAppendixListingCaption}[2]{%
  \par\addvspace{7pt}%
  \refstepcounter{controllerAppendixListing}\label{#2}%
  {\centering\small Listing~\arabic{controllerAppendixListing}. #1\par}%
  \nopagebreak[4]%
}
\section{Four-Slice Prompts and Controller Code}
\label{app:four-slice-prompts}
This appendix presents the shared task prompt and mutation template
for the four-slice simulation, followed by an example evolved controller.
The system prompt remains fixed. Initial generation uses the request
in Listing~\ref{lst:four-slice-initial}; mutation replaces that request
with the parent program and feedback in
Listing~\ref{lst:four-slice-mutation}.

\subsection{Task Prompt}
The prompt supplies the capacity constraints and states that some
throughput targets can become infeasible.
\controllerAppendixListingCaption{Shared four-slice task prompt (system message).}{lst:four-slice-system}
\begin{lstlisting}[style=controllerAppendixPrompt]
You are designing a PRB-allocation controller for a shared radio pool with four GUARANTEED network slices (G1, G2, G3, G4) and one BEST-EFFORT (saturated) slice. Each guaranteed slice k has a fixed weight w_k and a per-slice throughput target T_k (Mbps): (w, T) = (1.00, 44), (0.90, 46), (0.80, 48), (0.70, 50) for G1, G2, G3, G4 respectively.
Write a pure Python function EXACTLY named `controller` with signature:
    def controller(history, state, sla, predictors):
It returns a LIST of 4 numbers (s1, s2, s3, s4): the PRB shares to give the four guaranteed slices. The harness clamps negative entries to 0 and, if the four shares sum to more than 70, scales them down to sum 70; the best-effort slice receives the rest of the pool (the pool is 100, so the best-effort slice always receives at least 30). You choose the shares for the four guaranteed slices.
Objects:
  history.slices : list of 4 per-slice rolling histories (index 0 = G1 ... index 3 = G4).
    history.slices[k].last() -> most recent sample or None; sample fields: .cqi (slice k's channel quality, 5..15), .prb_share (its last effective share), .dl_bytes (its bytes this tick; Mbps = dl_bytes*8/1e6). history.slices[k].latest_mbps() returns its last delivered Mbps.
  state : persistent dict-like scratch (state.get(key,default), state.set(key,val)).
  sla   : sla.slices[k].weight (w_k), sla.slices[k].target_mbps (T_k), sla.pool_max (=100), sla.guaranteed_cap (=70, the maximum the four guaranteed shares may sum to), sla.n (=4), sla.best_effort (the best-effort SLA).
  predictors : ignore (may be empty).
A guaranteed slice's delivered throughput on a tick is min(its demand, rate_k * share_k), where rate_k is its Mbps per share and its demand stays at or above its target. SCORING - maximize the per-tick average of:
    score = sum_k [ w_k * attain_k - 0.5 * w_k * (1 - attain_k) ] + 0.6 * (best_effort_Mbps / CEIL) - 0.02 * churn
where the sum is over the four guaranteed slices, attain_k = 1 if slice k's delivered throughput >= T_k - 0.5 else 0, best_effort_Mbps is the best-effort slice's delivered throughput, CEIL = 239, and churn = the number of the (up to five) shares that changed from the previous tick. Your controller observes each guaranteed slice's CQI each tick. Design the controller to MAXIMISE the average score.
Return ONLY a fenced ```python code block with the function (helpers/constants ok). NO imports, no I/O.

Notes: (i) the four guaranteed channels vary independently over time; (ii) there are sustained states in which the pool cannot afford all four targets at once, so in those states some guaranteed targets go unmet; (iii) in the deepest degradations a slice's target cannot be met at any share; (iv) reported CQI has a floor - in a deep null a slice's CQI reads approximately 5.0, its delivered throughput per share drops near zero, and its target cannot be met at any share, while in a good channel CQI saturates near 15 and each share delivers more Mbps; (v) delivered throughput is capped by current demand, and each guaranteed slice's demand stays at or above its target.
\end{lstlisting}
\controllerAppendixListingCaption{Initial user message.}{lst:four-slice-initial}
\begin{lstlisting}[style=controllerAppendixPrompt]
Design the controller function for the four-guaranteed-slice system described above to maximise the average score. Handle empty history robustly. Return ONLY the fenced ```python code block.
\end{lstlisting}

\subsection{Mutation Template}
The fields below are filled with the parent program and its evaluation
results. Attainment is reported separately for the four guaranteed slices.
\controllerAppendixListingCaption{Mutation user-message template.}{lst:four-slice-mutation}
\begin{lstlisting}[style=controllerAppendixPrompt]
Here is a controller with worst-case fitness {fitness:+.4f}.

Diagnostics (means over evaluation): per-slice attainment={attainment}, violation%={violation:.1f}, best-effort Mbps={best_effort:.1f}, share-changes/episode={churn}.

```python
{parent controller source}
```

Produce an IMPROVED variant that scores higher on the average score. Return ONLY a fenced ```python code block.
\end{lstlisting}

\subsection{Controller Code}
The following example returns four allocation shares. Comments and
docstrings are omitted for readability; the executable code is unchanged.
The prompt requests no imports, while this implementation imports
\texttt{math} to use \texttt{math.ceil}.
\controllerAppendixListingCaption{An evolved controller for the four-slice simulation.}{lst:four-slice-controller}
\begin{lstlisting}[style=controllerAppendixCode]
import math
TAU = (9.0, 9.0, 11.0, 11.0)
def _rate_of_cqi(cqi):
    c = 5.0 if cqi < 5.0 else 15.0 if cqi > 15.0 else cqi
    return 0.30 + ((2.39 - 0.30) / (10.0 ** 1.578)) * (c - 5.0) ** 1.578
def controller(history, state, sla, predictors):
    if not state.get('prev_shares', None):
        state.set('prev_shares', [0, 0, 0, 0])
        state.set('attain_history', [[], [], [], []])
        state.set('cqi_history', [[], [], [], []])
    prev_shares = state.get('prev_shares')
    attain_history = state.get('attain_history')
    cqi_history = state.get('cqi_history')
    cqis = []
    rates = []
    targets = []
    weights = []
    recent_mbps = []
    for k in range(sla.n):
        h = history.slices[k].last()
        cqi = h.cqi if h is not None else 15.0
        cqis.append(cqi)
        rates.append(_rate_of_cqi(cqi))
        targets.append(sla.slices[k].target_mbps)
        weights.append(sla.slices[k].weight)
        recent_mbps.append(history.slices[k].latest_mbps() or 0.0)
        cqi_history[k].append(cqi)
        if len(cqi_history[k]) > 20:
            cqi_history[k].pop(0)
        attain = 1.0 if recent_mbps[k] >= targets[k] - 0.5 else 0.0
        attain_history[k].append(attain)
        if len(attain_history[k]) > 20:
            attain_history[k].pop(0)
    window = 5
    recent_attain = []
    for k in range(sla.n):
        if len(attain_history[k]) > 0:
            recent_rate = sum(attain_history[k][-window:]) / len(attain_history[k][-window:])
        else:
            recent_rate = 0.5
        recent_attain.append(recent_rate)
    adaptive_tau = list(TAU)
    for k in range(sla.n):
        if recent_attain[k] > 0.7:
            adaptive_tau[k] = TAU[k] - 0.5
        elif recent_attain[k] < 0.3:
            adaptive_tau[k] = TAU[k] + 0.5
    min_shares = []
    for k in range(sla.n):
        if rates[k] > 0.001:
            min_share = math.ceil(targets[k] / rates[k])
        else:
            min_share = sla.guaranteed_cap + 1
        min_shares.append(min_share)
    shares = [0, 0, 0, 0]
    viable = []
    for k in range(sla.n):
        if cqis[k] >= adaptive_tau[k] - 0.5 and min_shares[k] <= sla.guaranteed_cap:
            cqi_quality = (cqis[k] - 5.0) / 10.0
            momentum = 0.5 + 0.5 * recent_attain[k]
            viability = weights[k] * cqi_quality * momentum
            viable.append((viability, k))
    viable.sort(reverse=True)
    total_assigned = 0
    assigned_set = set()
    for _, k in viable:
        min_need = min_shares[k]
        if total_assigned + min_need <= sla.guaranteed_cap:
            shares[k] = min_need
            total_assigned += min_need
            assigned_set.add(k)
    remaining = sla.guaranteed_cap - total_assigned
    if remaining > 0 and assigned_set:
        assigned_list = list(assigned_set)
        weight_sum = sum(weights[k] for k in assigned_list)
        if weight_sum > 0:
            for k in assigned_list:
                extra = int(remaining * weights[k] / weight_sum)
                shares[k] += extra
                remaining -= extra
            if remaining > 0:
                best_k = max(assigned_list, key=lambda k: weights[k])
                shares[best_k] += remaining
    for k in range(sla.n):
        prev_share = prev_shares[k]
        curr_share = shares[k]
        if prev_share > 0 and curr_share == 0:
            if (weights[k] > 0.8 or recent_attain[k] > 0.4) and cqis[k] >= 7.0:
                retain_share = min(1, sla.guaranteed_cap - sum(shares))
                if retain_share > 0:
                    shares[k] = retain_share
    cqi_delta_threshold = 2.0
    for k in range(sla.n):
        prev_share = prev_shares[k]
        if prev_share > 0:
            if len(cqi_history[k]) >= 2:
                cqi_delta = abs(cqis[k] - cqi_history[k][-2])
            else:
                cqi_delta = 0
            if cqi_delta < cqi_delta_threshold:
                max_change = max(1, prev_share // 2)
                shares[k] = max(0, min(prev_share + max_change, shares[k]))
    total = sum(shares)
    if total > sla.guaranteed_cap:
        adjustable = [(shares[k] - min_shares[k], k) for k in range(sla.n) if shares[k] > 0]
        adjustable.sort(reverse=True)
        excess = total - sla.guaranteed_cap
        for _, k in adjustable:
            if excess <= 0:
                break
            reduction = min(excess, shares[k] - min_shares[k])
            shares[k] -= reduction
            excess -= reduction
    shares = [max(0, min(s, sla.guaranteed_cap)) for s in shares]
    total = sum(shares)
    if total > sla.guaranteed_cap:
        scale = sla.guaranteed_cap / max(total, 1)
        shares = [int(s * scale) for s in shares]
    state.set('prev_shares', shares[:])
    state.set('attain_history', attain_history)
    state.set('cqi_history', cqi_history)
    return shares
\end{lstlisting}

\end{document}